\documentstyle[12pt]{article}
\advance\oddsidemargin by -1in
\advance\evensidemargin by -1in
\advance\topmargin  by -0.5in
\input epsf.tex

\newcommand{\sx}{\sigma^{x}}
\newcommand{\sz}{\sigma^{z}}

\newcommand{\deff}{{\stackrel{{\rm def}}{=}}}
\newcommand{\rmv}{{\rm v}}
\newcommand{\CC}{{\bf C}}
\newcommand{\LL}{{\bf L}}
\newcommand{\ZZ}{{\bf Z}}

\title{Quantum codes on a lattice with boundary.\,\thanks{
This work has been supported, in part, by the Russian Foundation for
Fundamental Research (grant no.~96-01-01113).}}
\author{
Sergey~B.~Bravyi$\,^\dag$ and Alexei~Yu.~Kitaev$\,^\ddag$\medskip\\
$^\dag$ {\it L.~D.~Landau Institute for Theoretical Physics},\\
{\it Kosygina St.~2, Moscow, 117940, Russia}\\
$^\ddag$ {\it California Institute of Technology, Pasadena, CA 91125, U.S.A.}\\
and\, {\it L.~D.~Landau Institute for Theoretical Physics.}}

\begin{document}
\maketitle 

\begin{abstract}
A new type of local-check additive quantum code is presented. Qubits are
associated with edges of a 2-dimensional lattice whereas the stabilizer
operators correspond to the faces and the vertices. The boundary of the
lattice consists of alternating pieces with two different types of boundary
conditions. Logical operators are described in terms of relative homology
groups.
\end{abstract}
 
Since Shor's discovery of the quantum error correcting codes~\cite{Shor}, a
large number of examples have been constructed. Most of them belong to the
class of additive codes~\cite{additive}. More specifically, codewords of an
additive code form a common eigenspace of several commuting {\em stabilizer
operators}, each of which is a product of Pauli matrices acting on different
qubits. A peculiar property of toric codes~\cite{ercor,quco,anyons} is that
the stabilizer operators are {\em local}: each of them involves only 4 qubits,
each qubit is involved only in 4 stabilizer operators, while the code distance
goes to infinity. (The number 4 is not a matter of principle; it could be any
constant). Furthermore, this locality is geometric while the codeword subspace
and error correction properties are related to the {\em topology} of the
torus. Operators acting on codewords are associated with 1-dimensional
homology and cohomology classes of the torus (with $\ZZ_2$
coefficients). Similar codes can be defined for lattices on an arbitrary
closed 2-D surface.  In this paper we extend this definition to surfaces with
boundary. A similar construction has been proposed by M.~Freedman and
D.~Meyer~\cite{FM}.

Let us briefly recall the definition and the properties of the toric codes.

In a toric code, qubits are associated with edges of an $n\times n$ square
lattice on the torus $T^2$. To each vertex $s$ and each face $p$ we assign a
stabilizer operator of the form:
\begin{equation}
\label{st-op}
A_s\,= \prod_{j \in {\rm star}(s)} \sx_j \ , \qquad B_p\,=
\prod_{j \in {\rm boundary}(p)} \sz_j \,.
\end{equation}
(Note the dependencies between the stabilizer operators:
$\prod_{s}A_s=\prod_{p}B_p=1$).  A codeword is a vector $|\xi\rangle$ which
satisfies the following conditions
\begin{equation} \label{codeword-subspace}
A_s |\xi\rangle\,=\,|\xi\rangle\,, \qquad B_p |\xi\rangle\,=\,|\xi\rangle
\qquad \mbox{for all}\ s,p\,.
\end{equation}

The codeword subspace $C$ is 4-dimensional, so it can be identified with the
Hilbert space of two {\em logical qubits}. This identification goes through
the algebra $\LL(C)$ of operators acting on $C$ which we call {\em logical
operators}. (They are also called informational operators~\cite{quco}). Any
logical operator can be extended to the large Hilbert space
$({\CC^2})^{\otimes 2n^2}$ of the physical qubits (which are associated to the
edges of the lattice). Such an extension is not unique but it can be chosen so
that to commute with all $A_s$ and $B_p$.  Operators with this property form
an algebra $\cal G$. In order to get the algebra $\LL(C)$, we take into
account that $A_s$ and $B_p$ act on $C$ as the identity operator and thus
should be identified with the identity. In a rigorous language, $\LL(C)$ is
the quotient of $\cal G$ by the ideal generated by $A_s-1$ and $B_p-1$. (This
applies to any additive code).

To be more particular, consider an operator of the form
\begin{equation}
\label{oper}
Y(c,c^*) \,\deff\, \prod_{i \in c} \sz_i \prod_{j \in c^*} \sx_j \, ,
\end{equation}
where $c$ is a 1-cycle with $\ZZ_2$ coefficients, $c^*$ is a 1-cycle on the
dual lattice.  (The edges of the original and the dual lattice are in 1-to-1
correspondence, so the two lattices share the qubits). The operator $Y(c,c^*)$
commutes with every stabilizer operator and thus maps the codeword subspace
$C$ to itself. This map depends only upon the homology classes of the cycles
$c$ and $c^*$, so we can denote it by $Y([c],[c^*])$. (In terms of the general
construction from the previous paragraph, the transition from cycles to
homology classes corresponds to quotienning by $A_s-1$ and $B_p-1$). Thus the
operators $Y(c,c^*)$ form a {\em linear basis} of the algebra $\cal G$ whereas
$Y([c],[c^*])$ form a {\em linear basis} of $\LL(C)$.~\footnote{ Note that
operators form linear spaces over $\CC$ while the cycles and homology classes
have a $\ZZ_2$ additive structure. There is nothing wrong here because
addition of cycles corresponds to operator multiplication.}  Let $[c_1]$,
$[c_2]$ be some basis elements of the group $H_1\left(T^2,\ZZ_2\right)$, and
$[c^*_1]$, $[c^*_2]$ form the dual basis of $H^1 \left(T^2,\ZZ_2\right)$. We
can represent these homology and cohomology classes by cycles $c_1$, $c_2$,
$c^*_1$, $c^*_2$ on the original and the dual lattice, respectively. The
corresponding logical operators $Y^x_1 =Y(0,[c^*_1])$, $Y^x_2 =Y(0,[c^*_2])$,
$Y^z_1 =Y([c_1],0)$, $Y^z_2=Y([c_2],0)$ are {\em generators} of the algebra
$\LL(C)$. They have the same commutation relations as $\sx_1$, $\sx_2$,
$\sz_1$, $\sz_2$, so we can map ones to the others. This way we establish an
isomorphism between the algebra $\LL(C)$ and the algebra
$\LL(\CC^2\otimes\CC^2)$, whence the correspondence between the codewords and
quantum states of two qubits.

This construction will be our starting point. Instead of dealing with toric
lattices, we consider a finite square lattice on the plane. A new feature
arising here is a boundary. Generally, the boundary can be of two types, see
Fig.~1. We will call them an $x$-boundary and a $z$-boundary.

\begin{figure}[h]
\epsfysize 30mm
\centerline{\epsfbox{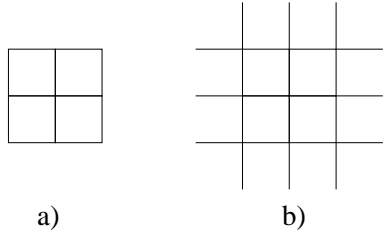}}
\caption{\small Square lattices with {\bf a)} $z$-boundary and {\bf b)}
$x$-boundary.}
\end{figure}
 
The simplest example of a boundary code can be built on the lattice having two
pieces of $x$-boundary and two pieces of $z$-boundary, in alternating order
(see Fig.~2). Under a suitable convention, an $n\times m$ lattice has $nm$
vertical edges and $(n+1)(m+1)$ horizontal edges, so the code has $2nm+n+m+1$
qubits. The stabilizer operators are very similar to ones in the toric
code. The definitions (\ref{st-op}), (\ref{codeword-subspace}) remain
essentially the same, but we must specify what are the faces and the
vertices. If a face $p$ is such that all its boundary edges are present
(e.~g.\ the face $p_2$ in the Fig.~2) then the operator $B_p$ is well defined
by~(\ref{st-op}). There are also incomplete faces lacking one edge, e.~g.\ the
face $p_1$ in Fig.~2. We still assign a stabilizer operator to such a face
according to~(\ref{st-op}), with $\mbox{boundary($p$)}$ containing all
existing boundary edges of the face $p$. Thus there are $n(m+1)$ face
stabilizer operators. Similarly, we assign $(n+1)m$ stabilizer operators to
all vertices with 4 or 3 incoming edges. (Free ends of edges do not bear
stabilizer operators). All the stabilizer operators are independent.
\begin{figure}[h]
\epsfysize=70mm
\centerline{\epsffile{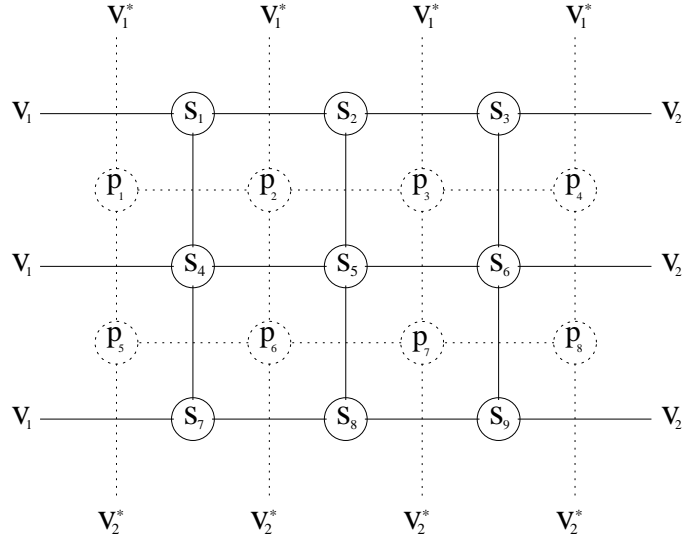}}
\caption{\small A $2\times 3$ lattice with
two pieces of $x$-boundary and two pieces of $z$-boundary.
The free ends labeled by the same letter could be identified.}
\end{figure}

Here is a complete list of stabilizer operators for the lattice shown
in Fig.~2:
$ A_{s_1} = \sx_{\rmv_1 s_1} \sx_{s_1 s_4} \sx_{s_1 s_2}$,
$ A_{s_2} = \sx_{s_1 s_2} \sx_{s_2 s_5} \sx_{s_2 s_3}$,
$ A_{s_3} = \sx_{s_2 s_3} \sx_{s_3 s_6} \sx_{s_3 \rmv_2}$,
$ A_{s_4} = \sx_{\rmv_1 s_4} \sx_{s_1 s_4} \sx_{s_4 s_5} \sx_{s_4 s_7}$,
$ A_{s_5} = \sx_{s_4 s_5} \sx_{s_2 s_5} \sx_{s_5 s_6} \sx_{s_5 s_8}$,
$ A_{s_6} = \sx_{s_5 s_6} \sx_{s_3 s_6} \sx_{s_6 \rmv_2} \sx_{s_6 s_9}$,
$ A_{s_7} = \sx_{\rmv_1 s_7} \sx_{s_4 s_7} \sx_{s_7 s_8}$,
$ A_{s_8} = \sx_{s_7 s_8} \sx_{s_5 s_8} \sx_{s_8 s_9}$,
$ A_{s_9} = \sx_{s_8 s_9} \sx_{s_6 s_9} \sx_{s_9 \rmv_2}$ and
$B_{p_1} = \sz_{\rmv_1 s_1} \sz_{s_1 s_4} \sz_{\rmv_1 s_4}$,
$B_{p_2} = \sz_{s_1 s_2} \sz_{s_2 s_5} \sz_{s_4 s_5} \sz_{s_1 s_4}$,
$B_{p_3} = \sz_{s_2 s_3} \sz_{s_3 s_6} \sz_{s_5 s_6} \sz_{s_2 s_5}$,
$B_{p_4} = \sz_{s_3 \rmv_2} \sz_{s_3 s_6} \sz_{s_6 \rmv_2}$,
$B_{p_5} = \sz_{\rmv_1 s_4} \sz_{s_4 s_7} \sz_{s_7 \rmv_1}$,
$B_{p_6} = \sz_{s_4 s_5} \sz_{s_5 s_8} \sz_{s_7 s_8} \sz_{s_4 s_7}$,
$B_{p_7} = \sz_{s_5 s_6} \sz_{s_6 s_9} \sz_{s_8 s_9} \sz_{s_5 s_8}$,
$B_{p_8} = \sz_{s_6 \rmv_2} \sz_{s_6 s_9} \sz_{s_9 \rmv_2}$.

The dimensionality of the codeword subspace can be found by a simple counting
argument.  There are $2nm+n+m+1$ qubits and $2nm+n+m$ independent stabilizer
operators which leave us with $(2nm+n+m+1)-(2nm+n+m)=1$ degrees of freedom,
i.~e.\ only one logical qubit can be encoded. Thus the codeword subspace $C$
is 2-dimensional. Let us find the logical operators acting on it. Firstly, we
are to characterize the algebra $\cal G$ of operators commuting with all the
stabilizer operators. Then we will find $\LL(C)$ by taking a quotient.

Let us denote the lattice and the dual lattice by $L$ and $L^*$,
respectively. Both lattices have boundaries which are, by definition, formed,
by free ends of edges. (Recall that the free ends are exactly the vertices
which do not bear stabilizer operators). Note that the $x$-boundary belongs to
the lattice $L$ while the $z$-boundary belongs to $L^*$. From now on, we
denote these two boundaries by $V$ and $V^*$, correspondingly. In Fig.~2, $V$
includes the free ends denoted by $V_1$ and $V_2$, whereas $V^*$ is
represented by $V^*_1$ and $V^*_2$. (It does not matter whether we identify
the free ends or consider them as distinct vertices).

A linear basis of $\cal G$ is given by eq.~(\ref{oper}), where $c$ is a
relative 1-cycle (with $\ZZ_2$ coefficients) on the lattice $L$, and $c^*$ is
a relative 1-cycle on the lattice $L^*$. By definition, a {\em relative
1-cycle} on a lattice is a 1-chain $c$ whose boundary $\partial c$ is
contained in the boundary of the lattice. Equivalently, a relative 1-cycle is
an ordinary (or absolute) 1-cycle on a lattice obtained from the original one
by gluing all the free ends together. (To prove that the operators $Y(c,c^*)$
actually make up a linear basis of $\cal G$, expand a generic linear operator
into products of Pauli matrices and try to commute with $A_s$ and $B_p$).

The action of $Y(c,c^*)$ on the codeword subspace $C$ depends only upon the
relative homology classes $[c]\in H_1(L,V,\ZZ_2)$ and $[c^*]\in
H_1(L^*,V^*,\ZZ_2)=H^1(L,V,\ZZ_2)$. Thus we arrive to the group 
$E=H_1(L,V,\ZZ_2)\oplus H_1(L^*,V^*,\ZZ_2)$. The operators $Y([c],[c^*])$\,
(where $([c],[c^*])\in E$) form a linear basis of the algebra $\LL(C)$.

Consider some relative cycle $c_{12}$ starting at $V_1$ and ending at
$V_2$, and some relative cycle $c^*_{12}$ starting at $V^*_1$ and ending
at $V^*_{2}$, see Fig.~3.  The operators $Y^z=Y([c_{12}],0)$ and
$Y^x=Y(0,[c^*_{12}])$ generate the algebra of logical operators. Since these
two generators anti-commute, we can interpret them as the action of $\sz$ and
$\sx$ on the logical qubit.

\begin{figure}[h]
\epsfysize=35mm
\centerline{\epsfbox{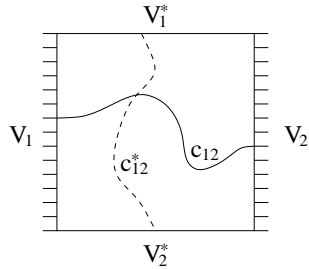}}
\caption{\small The nontrivial relative homology class $[c_{12}]\in
H_1(L,V,\ZZ_2)$ is shown by a solid line. The nontrivial element
$[c^*_{12}]\in H_1(L^*,V^*,\ZZ_2)$ is shown by a dashed line.}
\end{figure}

Let us find the distance of the code we have constructed. By definition, the
distance of a code is the minimal size of an error which can not be detected
by syndrome measurement but still affects the codeword subspace $C$. A general
error is just an operator on the Hilbert space of physical qubits. An error
$X$ is undetectable if it commutes with the stabilizer operators, i.e.\
belongs to $\cal G$. As we know, such operators $X$ are linear combinations of
$Y(c,c^*)$. An operator $Y(c,c^*)$ affects the codeword subspace if at least
one of the relative cycles $c$ and $c^*$ is nontrivial. Thus, the code
distance is the length of a shortest path which connects two pieces of
boundary of the same type. For an $n\times m$ lattice, this number equals
$d=\min\{n+1,\,m+1\}$.  The code protects against
$\left\lfloor\frac{d-1}{2}\right\rfloor$ errors.

A similar code can be defined for any pair of mutually dual lattices with
boundary. They need not be square lattices; each vertex can have any number of
neighbors, and each face can be of arbitrary size. We omit formal definition
here. Topologically, the pair of lattices corresponds to a surface $Q$ with
boundary split into pieces of two types, $x$ and $z$. The two parts of the
boundary will be denoted by $V$ and $V^*$, respectively. If we draw the
lattices on this surface, the free ends of the first lattice should be
attached to $V$ whereas the free ends of the dual lattice should be attached
to the $V^*$. The above arguments work perfectly in this general case. The
basis logical operators $Y([c],[c^*])$ correspond to relative homology classes
$[c]\in H_1(Q,V,\ZZ_2)$ and $[c^*]\in
H_1(Q,V^*,\ZZ_2)=H^1(Q,V,\ZZ_2)$. Hence the number of logical qubits is\,
$m=\dim H_1(Q,V,\ZZ_2)=\dim H_1(Q,V^*,\ZZ_2)$. The code distance is
\begin{equation}
d\, =\, \min{ \left\{ 
 \min_{[c] \ne 0}{ |{\it sup}p(c)|},\, \min_{ [c^*] \ne 0}{|{\it supp}(c^*)|}
\right\} }\, ,
\end{equation}
where $c$ and $c^*$ consist of edges of the corresponding lattices.

Let us consider the case where the surface $Q$ is a disk with $k$ pieces of
$x$-boundary (labeled as $V_i$) and $k$ pieces of $z$-boundary (labeled as
$V^*_i$), see Fig.~4. Obviously, $\dim H_1(Q,V,\ZZ_2)=k-1$, hence $k-1$
qubits can be encoded. A particular encoding can be specified if we select a
basis of $H_1(Q,V,\ZZ_2)$ and the dual basis of
$H_1(Q,V^*,\ZZ_2)=H^1(Q,V,\ZZ_2)$. For example, we can choose the operators
\begin{equation}
Y^z_i\,=\,Y([c_i],0)\ ,\quad\ Y^x_i \,=\,Y(0,[c^*_i])\ ,\quad\ i=1 \ldots k-1
\end{equation}
to represent the action of $\sz_i$ and $\sx_i$ on the logical qubits. (Here
$c_i$ is a path which connects $V_i$ with $V_{i+1}$, whereas $c^*_i$ connects
$V^*_i$ with $V^*_k$).

\begin{figure}[h]
\epsfysize=50mm
\centerline{\epsfbox{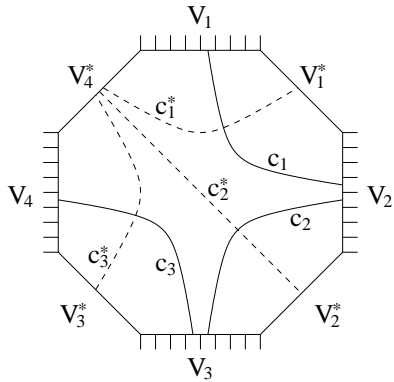}}
\caption{\small A lattice with $4+4$ pieces of boundary. Solid and dashed
lines represent the relative cycles $c_i$ and $c^*_i$ which correspond to the
logical operators $Y^z_i$ and $Y^x_i$, respectively.}
\end{figure}        

Finally, we try to explain the physical meaning of the two types of boundary
in terms of the topological quantum order (TQO)\footnote{ The term
``topological quantum order'' means nontrivial topological properties of the
ground state, nothing more specific.} and anyonic excitations in the bulk
system~\cite{anyons}. (Now we replace a code by a Hamiltonian).  Why only two
types of boundary conditions? Can one invent a combination of them? The answer
is ``No'', provided the boundary is {\em rigid}, i.e. does not carry gapless
excitations. A proof will be published elsewhere~\cite{stability}; now we only
want to give the idea.

The TQO in the bulk system is characterized by braiding and fusion properties
of anyons.  There are four sectors (i.e.\ fundamental particle types): the
vacuum sector (no particle), an ``electric charge'' (which lives on vertices),
a ``magnetic charge'' (which lives on the faces), and a combination of
both~\cite{anyons}. These sectors are stable with respect to weak generic
perturbations of the Hamiltonian. The stability can be explained by the the
nontrivial braiding properties of anyons. Indeed, an ``electric'' or
``magnetic'' charge can not simply disappear because that would change the
Berry phase of another particle moving around the charge at large
distance. (Note that that there is no non-topological long-range interaction
between the particles because all excitations in the system have energy
gap). However, an ``electric charge'' can disappear at the $x$-boundary, and a
``magnetic charge'' can disappear at the $z$-boundary. So, the bulk TQO is
unstable near the boundary. The two types of rigid boundary are just two
possible ways to resolve this instability. That is, there are two types of
{\em stable} boundary TQO consistent with the bulk TQO. One can prove that
these two types are the only possible ones. In particular, the combination of
an ``electric charge'' and a ``magnetic charge'' can not disappear at a rigid
boundary because this particle is a fermion. Note that a single ``electric''
or ``magnetic'' charge is a boson (with respect to itself).
\bigskip

\noindent{\bf Acknowledgements.} One of us (A.~K.) thanks Michael Freedman and
David Meyer for a valuable discussion which has helped to establish an
isomorphism between our construction and the construction in ref.~\cite{FM}.

\end{document}